\documentclass[aps,prb,preprint,floatfix,superscriptaddress,showpacs]{revtex4-1}

\usepackage{graphicx}
\usepackage{dcolumn}
\usepackage{amsmath,bm}
\usepackage{color}%
\usepackage{multirow}
\usepackage{array} 
\usepackage{xr}
\usepackage{makecell}

\makeatletter
\newcommand*{\addFileDependency}[1]{
  \typeout{(#1)}
  \@addtofilelist{#1}
  \IfFileExists{#1}{}{\typeout{No file #1.}}
}
\makeatother
 
\newcommand*{\myexternaldocument}[1]{%
    \externaldocument{#1}%
    \addFileDependency{#1.tex}%
    \addFileDependency{#1.aux}%
}

\myexternaldocument{si}

\begin{document}


\title{Giant piezoelectricity in group IV monochalcogenides with ferroelectric AA layer stacking}

\author{Seungjun Lee}
\affiliation{Department of Electrical and Computer Engineering, University of Minnesota, Minneapolis, Minnesota 55455, USA}

\author{Hyeong-Ryul Kim}
\affiliation{Department of Physics and Research Institute for Basic Sciences, Kyung Hee University, Seoul 02447, Korea}

\author{Wei Jiang}
\affiliation{Department of Electrical and Computer Engineering, University of Minnesota, Minneapolis, Minnesota 55455, USA}

\author{Young-Kyun Kwon}\email{ykkwon@khu.ac.kr}
\affiliation{Department of Physics and Research Institute for Basic Sciences, Kyung Hee University, Seoul 02447, Korea}
\affiliation{Department of Information Display, Kyung Hee University, Seoul 02447, Korea}

\author{Tony Low}\email{tlow@umn.edu}
\affiliation{Department of Electrical and Computer Engineering, University of Minnesota, Minneapolis, Minnesota 55455, USA}

\date{\today}
\begin{abstract}
%
%
%
The piezoelectricity of group IV monochalcogenides (MXs, with M = Ge, Sn and X = S, Se) has attracted much attention due to their substantially higher piezoelectric coefficients compared to other 2D materials. However, with increasing layer number, their piezoelectricity rapidly disappears due to the antiferroelectric stacking order, severely limiting their practical applications. Using first-principles calculations, we investigated the piezoelectricity of MXs with the ferroelectric AA stacking configuration, which has recently been stabilized in experiments. We found that AA-stacked MXs have a ferroelectric ground state with the smallest lattice constant among other stacking configurations, resulting in a giant piezoelectric coefficient, which is the first demonstration of a strategy where the piezoelectric coefficients can increase with the number of layers. This can be attributed to a strong negative correlation between the lattice constant along the armchair direction and the piezoelectric coefficient, and spontaneous compressive strain stabilized in ferroelectric AA stacking configuration.
\end{abstract}

\maketitle


\section{Introduction}
\label{Introduction}
An interesting connection between mechanical energy and electricity in piezoelectric materials makes them highly valuable in various applications, such as energy harvesting, sensors, and optoelectronic devices.~\cite{wang2006piezoelectric,wang2007nanopiezotronics,bao2021piezophototronic}
Moreover, two-dimensional piezoelectric materials (2DPMs) are of practical interest compared to their bulk counterparts because they are robust against depolarization fields, enabling a reduction of device size.~\cite{hinchet2018piezoelectric,zhang2021piezotronics,jin2023emerging}
Therefore, many efforts contributed in the last decade have led to the discovery of various types of monolayer or one-layer (1L) piezoelectric materials such as 
hexagonal (h-) group III-IV materials,~\cite{ares2020piezoelectricity} 
h-group II oxides,~\cite{blonsky2015ab} 
H-phase transition metal dichalcogenides (TMDs),~\cite{zhu2015observation,alyoruk2015promising} 
In$_2$Se$_3$,~\cite{zhou2017out,xue2018multidirection} 
CuInP$_2$S$_6$,~\cite{liu2016room,brehm2020tunable} 
and group IV monochalcogenides (MXs, with M = Ge, Sn and X = S, Se).~\cite{Fei2015,gomes2015enhanced,wang2017two,higashitarumizu2020purely,xu2022van}
Among the various candidates, MXs, in particular, have received much attention due to their piezoelectric coefficients, which surpass those of other two-dimensional materials by orders of magnitude.

Despite these promising advancements, the practical utilization of 2DPMs has been limited because ground-state multilayer stacking configurations often lose their piezoelectric response.~\cite{li2013probing,zhu2015observation}
The challenge arises from the interlayer dipole-dipole interactions, which naturally favor an antiparallel order, resulting in the restoration of spatial inversion symmetry.
Nevertheless, recent advances in materials handling and various growth techniques have paved the way for the experimental stabilization of metastable stacking configurations,~\cite{yoo2019atomic,higashitarumizu2020purely}
resulting in novel ferroelectric (FE) order in 2DPMs.
For example, interlayer twisting~\cite{woods2021charge,yasuda2021stacking} or sliding~\cite{xu2022van,meng2022sliding,deb2022cumulative} give rise to additional out-of-plane FE order, which is not possible in a 1L or minimum energy stacking configuration.
However, despite the fact that in-plane ferroelectricity is also crucial for piezoelectric response, there is limited understanding of the effect of interlayer interactions on the in-plane FE order and the corresponding piezoelectric response.

Through first-principles calculation based on density functional theory (DFT), we investigated the stacking-dependent piezoelectric responses of MXs. 
We first carefully evaluated the piezoelectric coefficients of monolayer MX to understand the large variations in their piezoelectric coefficients reported previously.
We found that an in-plane lattice constant (LC) along the armchair direction primarily controls the piezoelectric coefficients of MX due to the unique puckered structures.
We further revealed that, the choice of van der Waals (vdW) correction is particularly crucial for MXs, as it significantly affects their in-plane lattice constants not only in multilayer cases but also in the 1L limit, which is not expected for other 2D materials.
Therefore, we investigated the piezoelectric coefficients of AA and AC stacked MX with carefully chosen vdW corrections. We found a significant enhancement of the piezoelectric coefficients in AA stacked MX regardless of the choice of vdW corrections. The physical origin of this enhancement is a spontaneous compressive strain emerging in the AA stacking.
Our results provide a deeper understanding of the piezoelectric responses of MXs and pave the way for a novel approach to optimize the piezoelectricity in 2DPMs through stacking configuration.

\section{Computational details}
\label{Computational}

The first-principles DFT calculation~\cite{Kohn1965} was carried out using Vienna \textit{ab initio} simulation package (VASP).~\cite{Kresse1996} We employed the plane wave basis to expand the electronic wavefunctions with a kinetic energy cutoff of 500~eV. 
The projector-augmented wave pseudopotentials~\cite{{Blochl1994},{Kresse1999}} were used for the valence electrons, and the exchange-correlation (XC) functional was treated within the generalized gradient approximation of Perdew–Burke–Ernzerhof (PBE).~\cite{Perdew1996}
A sufficiently large vacuum region ($>20~\textrm{\AA}$) was included in the unitcell to avoid any spurious interlayer interactions.
The atomic basis was carefully relaxed until the Helmann-Feynman force acting on every atom was smaller than 0.001~eV/$\textrm{\AA}$, which is very crucial to get a converged piezoelectric coefficient.
The Monkhorst-Pack 21$\times$21$\times$1 $k$-mesh was used to sample the Brillouin zone.
To comprehensively investigate the effects of interlayer interactions, we used various types of vdW interactions including, simple Grimme methods,~\cite{grimme-d2,grimme-d3} as well as various non-local vdW-DF functionals.~\cite{vdwdf,vdwdf2,optB88,revvdwdf2}

The spontaneous polarization, $P$, of various MXs were calculated in the framework of the modern theory of polarization based on the Berry phase.~\cite{spaldin2012beginner} 
Then, the planar elastic stiffness coefficients $C_{ij}$ and piezoelectric strain coefficients $e_{ijk}$ were evaluated as~\cite{Fei2015}
\begin{equation}
C_{ij}=\frac{1}{A_0}\frac{{\partial}^2U}{\partial {\varepsilon}_{ii}\partial {\varepsilon}_{jj}},
\end{equation}
\begin{equation}
e_{ijk}=\frac{\partial P_i}{\partial {\varepsilon}_{jk}}, \\
\end{equation}
where ${\varepsilon}$, $U$, and $A_0$ are the strain, the total energy and the unitcell area. 
The effective thickness of the 1L MXs was treated as half of the out-of-plane LC of their bulk counterparts, which was used to evaluate an effective bulk polarization.
Then, the piezoelectric stress coefficients $d_{ij}$ can be calculated as
\begin{equation}
d_{11}=\frac{e_{111}C_{22}-e_{122}C_{12}}{C_{11}C_{22}-C^2_{12}}, 
\end{equation}
\begin{equation}
d_{12}=\frac{e_{122}C_{11}-e_{111}C_{12}}{C_{11}C_{22}-C^2_{12}}.
\end{equation}


\section{Results and discussion} 
\label{Results}

\begin{figure}[t]
\includegraphics[width=0.8\columnwidth]{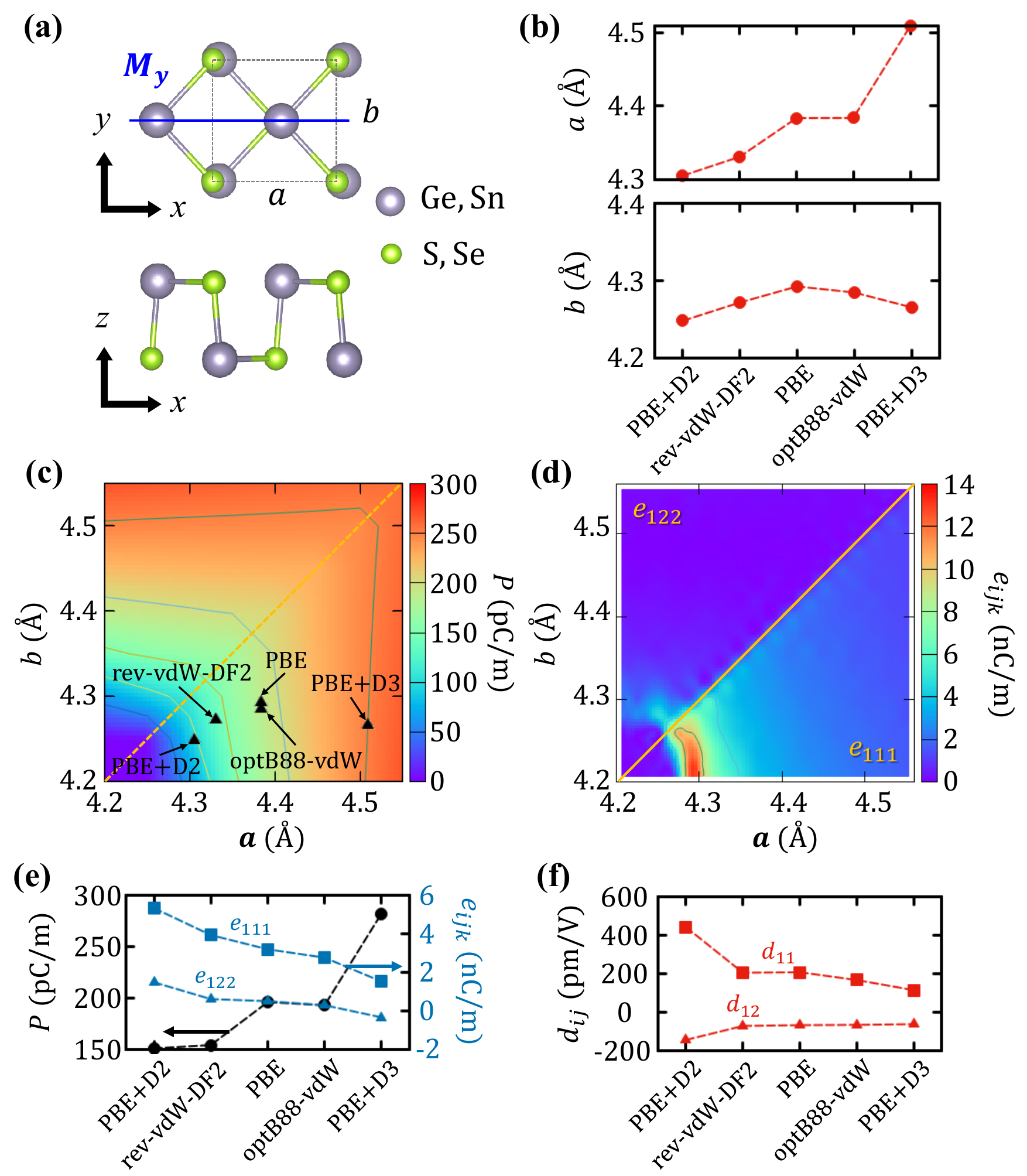}
\caption{
(a) Top and side views of a typical monolayer (1L) group IV monochalcogenides (MXs with M = Ge or Sn and X = S or Se) crystal structure. 
(b) Lattice constants, $a$ and $b$, of 1L SnSe obtained from different van der Waals (vdW). 
(c) Spontaneous polarization ${P}$ contour map of 1L SnSe, calculated by the PBE functional without vdW correction. Equilibrium lattice constants shown in (b) were overlaid for comparison. 
(d) 2D piezoelectric strain coefficient ($e_{ijk}$) contour map of 1L SnSe calculated by the derivative of (c) with strain along $a$. As a result, the upper left and lower right triangles represent $e_{111}$ and $e_{122}$, respectively. 
(e) ${P}$ and $e_{ijk}$, and (f) piezoelectric stress coefficients ($d_{ij}$) of 1L SnSe obtained from different vdW functionals.
\label{fig1}}
\end{figure}
\begin{table*}[t]
\caption {Lattice constants, $a$ and $b$, spontaneous polarization, $P$, piezoelectric strain coefficients, $e_{ijk}$, planar elastic stiffness coefficients, $C_{ij}$, and piezoelectric stress coefficients, $d_{ij}$ of 1L SnSe calculated by various functionals.
\label{table1}}
\begin{ruledtabular}
\begin{tabular}{ c c c c c c c c c c c } 
vdW  & \makecell{$a$ \\ (\AA)} & \makecell{$b$ \\ (\AA)} & \makecell{$P$ \\ (pC/m)} & \makecell{$e_{111}$  \\ (nC/m)} & \makecell{$e_{122}$  \\ (nC/m)} & \makecell{$C_{11}$  \\ (N/m)} & \makecell{$C_{22}$  \\ (N/m)} & \makecell{$C_{12}$  \\ (N/m)} & \makecell{$d_{11}$  \\ (pm/V)} & \makecell{$d_{12}$  \\ (pm/V)} \\
    \hline
none (PBE)	&	4.383 & 4.292	&	195.7	&	3.17	&	0.48	&	21.47	&	44.73	&	17.6	&	205.01	&	-69.98 \\
none (PBE)~\cite{Fei2015}	&	4.35 & 4.24	&		&	3.49	&	1.08	&	19.88	&	44.49	&	185.7	&	250.57	&	-80.31 \\
none (PBE)~\cite{guo2020small}	&	4.41 & 4.29  	&		&	2.81	&	0.52	&	23.06	&	42.82	&	18.89	&	175.32	&	-65.11 \\
none (PBE)~\cite{wang2017two}	&	 & 	&	181	&		&		&		&		&		&		&	 \\
D3	&	4.509 & 4.265	&	281.7	&	1.50	&	-0.36	&	25.47	&	42.06	&	21.02	&	112.29	&	-64.61 \\
D2	&	4.305 & 4.248	&	150.6	&	5.31	&	1.46	&	17.85	&	41.64	&	17.24	&	439.27	&	-146.81 \\
optB88	&	4.384 & 4.284	&	192.5	&	2.74	&	0.27	&	25.02	&	45.00	&	20.26	&	164.7	&	-68.15 \\
optB88~\cite{fang2016layer}	&	4.41 & 4.27	&		&	2.35	&	0.76	&	20	&	40.4	&	17	&	158.2	&	-47.7 \\
rev-vdW-DF2	&	4.331 & 4.271	&	153.8	&	3.91	&	0.59	&	26.61	&	47.6	&	20.19	&	202.8	&	-73.62 \\ 
\end{tabular}
\end{ruledtabular}
\label{table1}
\end{table*}

Figure~\ref{fig1}(a) shows the crystal structure of a typical 1L MX. We set $x$- and $y$-axes as in-plane directions with corresponding LCs of $a$ and $b$, and $z$-axis as out-of-plane direction. 
It has puckered atomic structures with a mirror plane of $M_{y}$ and belongs to $Pmn2_1$ space group.~\cite{gomes2015enhanced}
More importantly, its broken $M_{x}$ symmetry leads to a non-zero spontaneous polarization ${P}$ along $x$ or armchair direction.
Indeed, the spontaneous polarization of 1L MXs can be understood as a rearrangement of the atomic basis along the armchair direction. It has also been reported that the LC along the armchair direction plays a significant role in determining its ${P}$.~\cite{wang2017two} 
Therefore, we begin with our discussion on the LCs of 1L MXs using SnSe as an exemplary material and will carefully re-examine its piezoelectric coefficients.

Although the LCs of bulk MXs have been experimentally reported,~\cite{chattopadhyay1986neutron,PhysRevB.16.832,Murgatroyd2020} those of 1L MXs have not yet been done but are only available through DFT calculations. In most previous studies,~\cite{Fei2015,wang2017two,xu2022van} the LCs of 1L MXs have been obtained only with the PBE XC functional without vdW corrections, which were considered only in the multilayer case. This is the most common approach to determine the LCs of various 2D materials, since the effect of the vdW correction may be negligible in the 1L case. 
(See Fig.~S1 in Supplementary Information (SI) for more details.)
However, we found that the LC of 1L SnSe strongly depends on the choice of vdW correction, as shown in Fig.~\ref{fig1}(b).
For example, the LC ,$a$, obtained along the armchair direction ranges from 4.509~{\AA} (PBE+D3) to 4.305~{\AA} (PBE+D2), representing a deviation of almost 5\%. Such deviation is attributed to its puckered structure with a weak stiffness along the armchair direction. 

We found that a small change in the LCs of 1L MXs would cause a large variation in their piezoelectric coefficients. To understand this result, we calculated ${P}$ of 1L SnSe as a function of the two in-plane lattice constants {\textit{a}} and {\textit{b}} using PBE XC, which is shown as a contour map in Fig.~\ref{fig1}(c), where other calculated LCs using other functions are superimposed.
The contour map is symmetric with respect to the $b=a$ line with $C_{4z}$ rotational symmetry due to the degenerate ground states for $a>b$ and $a<b$.
As clearly visualized in Fig.~\ref{fig1}(c), $P$ is positively correlated with $a$, and the aspect ratio between $a$ and $b$ also has some secondary effects.
As the LC decreases, 1L SnSe loses its $P$ and becomes paraelectric.
Therefore, the vdW correction, which yields a smaller (larger) LC along the armchair direction, may underestimate (overestimate) ${P}$. For example, the $P$ value of 281.7~pC/m estimated with the PBE+D3 correction is almost twice the value of 150.6~pC/m estimated with the PBE+D2 correction.
Figure~\ref{fig1}(d) shows the contour map of the 2D piezoelectric strain coefficients, $e_{ijk}$, computed by taking the first derivative of $P$ shown in Fig.~\ref{fig1}(c) with respect to strain $\varepsilon_{jk}$, as explained in Computational details and Fig~S2
in SI. We found significantly stronger $e_{111}$ near the boundary between the paraelectric and FE phases due to the rapid emergence of non-zero ${P}$, and a small enhancement in $e_{122}$, although much smaller than $e_{111}$.
This result clearly indicates that a compressive strain along the armchair direction may play a major role in the enhancement of $e_{111}$ of SnSe. Simultaneously, it also implies that the fluctuation of the LC according to the choice of vdW functionals makes a large uncertainty in predicting its piezoelectric coefficients.

For a more systematic study, we evaluated the piezoelectric coefficients of 1L SnSe using various vdW functionals, as summarized in Fig.~\ref{fig1}(e, f), and Table~\ref{table1}.
Due to the strong correlation between ${P}$ and $a$, both the strain and stress piezoelectric coefficients ($e_{ijk}$ and $d_{ij}$) also strongly depend on the choice of functional.
For example, $e_{111}$ and $d_{11}$ range from from 5.31~nC/m to 1.50~nC/m, and from 439.27~pm/V to 112.29~pm/V, respectively.
As predicted above, the smaller $a$ leads to the higher $e_{ijk}$ and $d_{ij}$.
We further confirmed that our results are consistent with previously reported studies, and the observed variation in both $e_{ijk}$ and $d_{ij}$ is simply understood as the result of variation in the relaxed $a$.
Therefore, we emphasize that the choice of computational options is extremely important in the quantitative analysis and the vdW functional should be used consistently in both 1L and multilayer cases.

\begin{figure*}[t]
\includegraphics[width=\textwidth]{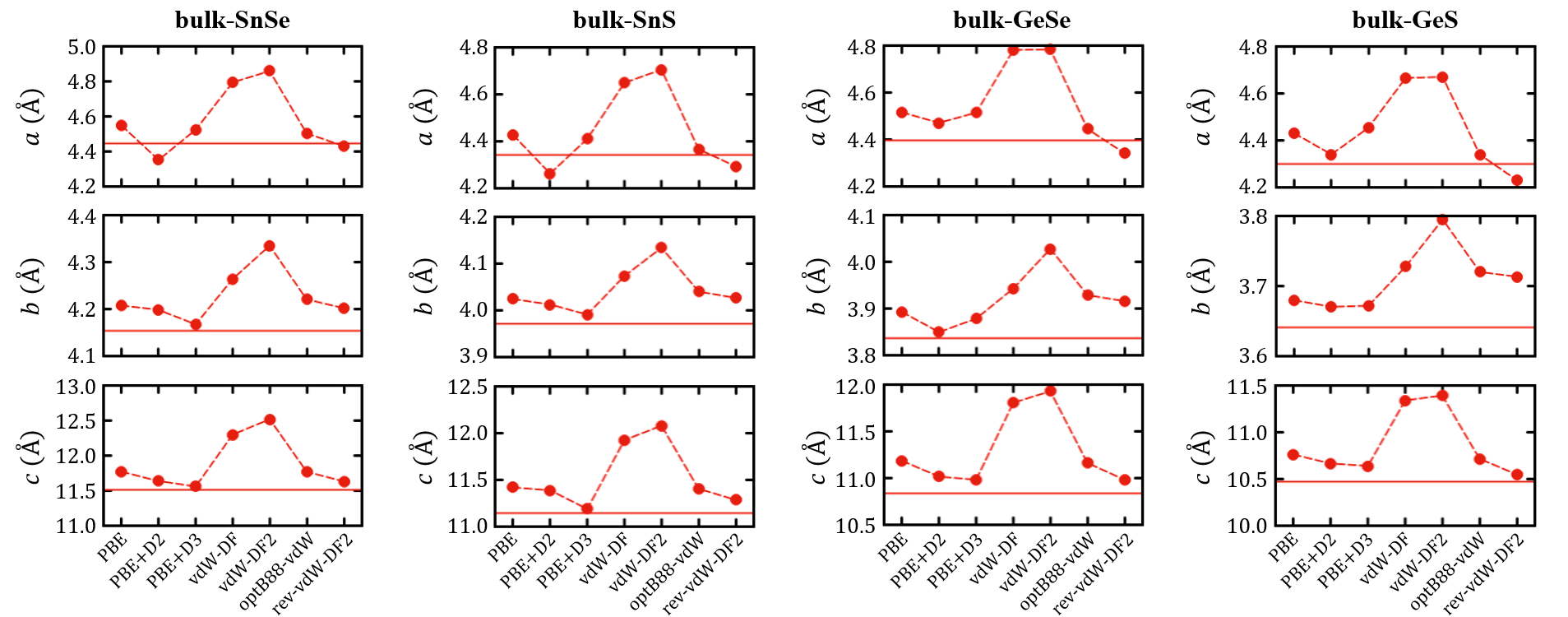}
\caption{Calculated lattice constants of bulk group IV monochalcogenides, obtained using various van der Waals corrections (red dots). Experimental values obtained from Ref. [\!\!\citenum{chattopadhyay1986neutron,PhysRevB.16.832,Murgatroyd2020}]
are also shown as red solid lines for comparison.
\label{fig2}}
\end{figure*}

A careful investigation in the previous section indicates that there is a significant uncertainty in determining the exact value of the piezoelectric coefficients of 1L MXs without experimental input of LCs.
Therefore, before investigating the piezoelectric coefficients of multilayer MXs, we calculated the LCs of bulk MXs, and then compared them with experimentally available values to validate which vdW correction is most appropriate among various vdW functionals.
Figure~\ref{fig2} shows the LCs of four different bulk MXs calculated with seven different functionals (red dots) together with their experimental values (red solid lines) for comparison.  
Among various functionals, the rev-vdW-DF2 method shows better agreement with experimental values than the other functionals.
(See also Figure~S3 and Table~S1 in SI)
Therefore, here, we primarily use the rev-vdW-DF2 vdW correction in investigating piezoelectric coefficients of 1L and multilayer MXs. 
For better understanding, we also repeated all calculations using the Grimme-D3 vdW corrections, with all results presented in SI.
Note that, all physical results are qualitatively consistent when using the same vdW functionals for both 1L and multilayer cases, regardless of which vdW functionals were used.
However, the conventional approach (i.e., vdW functional is only considered in multilayers) may lead to misinterpretation of the results, which will be discussed later. 

\begin{figure}[b]
\includegraphics[width=1.00\columnwidth]{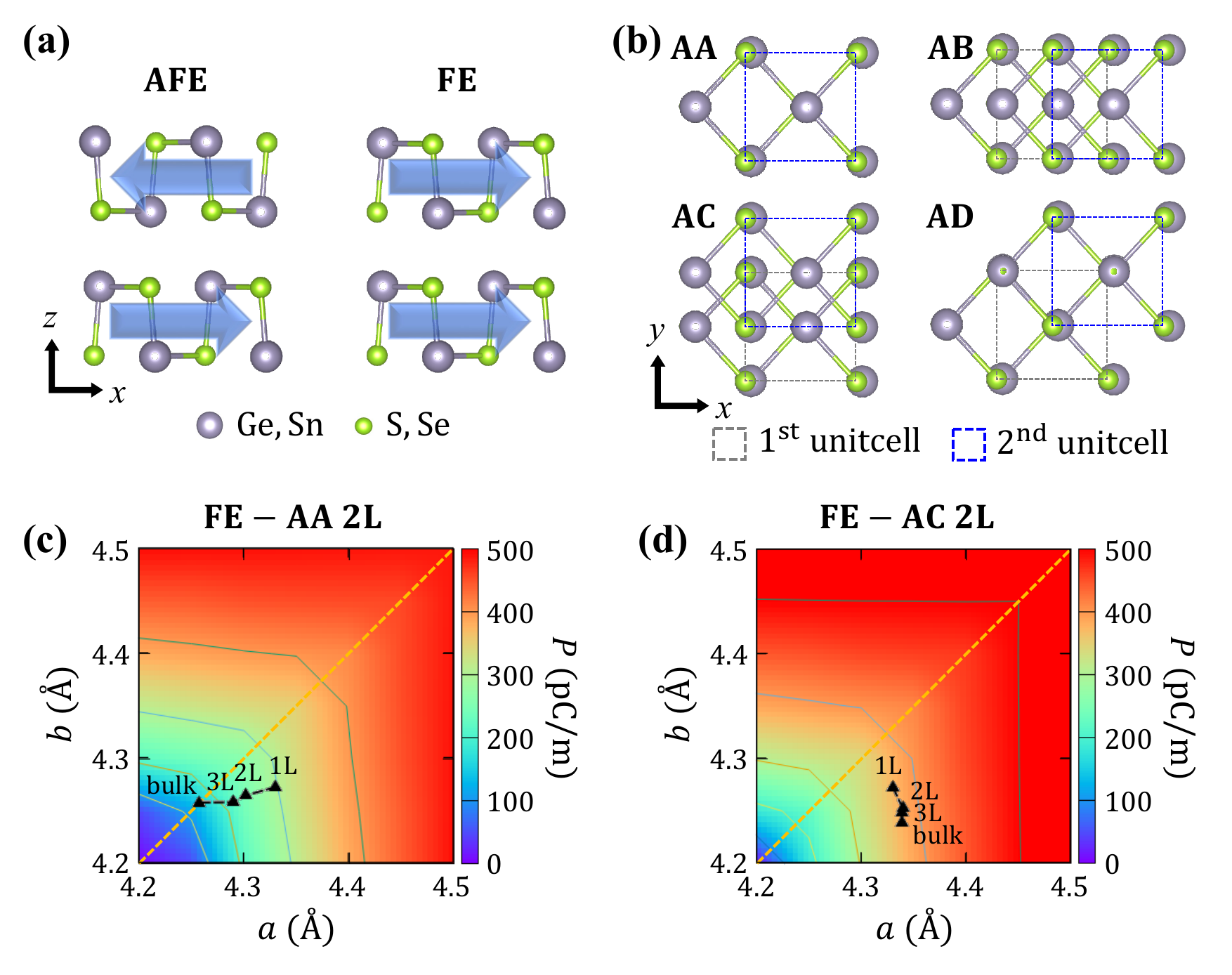}
\caption{(a) Side views of antiferroelectric (AFE) and ferroelectric (FE) stacking configurations of bilayer MX and (b) top views of four possible in-plane sliding configurations. In (a), blue arrows indicate the direction of the spontaneous polarization of each layer. In (b), grey and blue dashed rectangles visualize unitcells of each layer. 
(c) Contour maps of the spontaneous polarization $P$ of the FE 2L SnSe in (c) AA and (d) AC stacking configurations calculated by PBE with the rev-vdW-DF2 vdW correction. Equilibrium lattice constants of 1L, 2L, 3L, and bulk AA and AC SnSe are overlaid on (c) and (d), respectively. 
\label{fig3}}
\end{figure}

Figure~\ref{fig3}(a) and (b) show antiferroelectric (AFE) and FE stacking orders and four possible sliding configurations. 
Thus, there are a total of eight possible stacking configurations.
Although it is known that the ground state stacking configuration of MXs is the AFE AB stacking,~\cite{xu2022van,higashitarumizu2020purely} it has also been predicted that the FE order can be stabilized in both the AA and AC stacking configurations.~\cite{xu2022van} 
In addition, it has been experimentally confirmed that the AA-stacked FE SnS can exist on a mica substrate below a certain critical thickness.~\cite{higashitarumizu2020purely} Therefore, here, we first focus on 2L AA and AC stacked SnSe to understand the FE order in multilayer MXs.

Figure~\ref{fig3}(c) and (d) show the $P$ contour maps of FE 2L SnSe in AA and AC stacking configurations, respectively.
Both contour maps are nothing but those of their 1L counterparts shown in Fig.~\ref{fig1}(c) with twice larger $P$ values, implying that the ${P}$ of 2L SnSe is mainly due to the intralayer contribution, regardless of the stacking configuration.
Therefore, similar to the 1L case, the equilibrium LCs primarily determine their ${P}$.
The obtained LCs of multilayer SnSe were overlaid in Fig.~\ref{fig3}(c) and (d).
The LCs of 2L AA SnSe were evaluated to be $a=4.302$~{\AA} and $b=4.263$~{\AA}, whereas those of 2L AC phase are $a=4.340$~{\AA} and $b=4.251$~{\AA}. Especially we focus on change in $a$ along the armchair direction when stacked from 1L to 2L. As shown in Fig.~\ref{fig3}(c) and (d), where the LCs of SnSe with different number of layers are overlaid, AA stacking significantly reduces $a$, while AC stacking does not change $a$ considerably, compared to the 1L case.
Thus, 2L AA SnSe has a smaller ${P}$ and may have a larger $e_{ijk}$ than 2L AC SnSe.
It is worth noting that the reduction of LCs of 2L AA SnSe can be understood as a result of dipole-dipole interactions.
Since parallel dipole-dipole interaction is inherently unfavorable, relaxed structures can form in ways that favor either a decrease in ${P}$ or an increase in the ${P}$-${P}$ distance.
Notably, this result remains robust across different vdW functionals and is thus applicable to other MXs, emphasizing the generalizability of the observed phenomenon. 
(See Table~S2 in SI)

\begin{figure}[t]
\includegraphics[width=1.0\columnwidth]{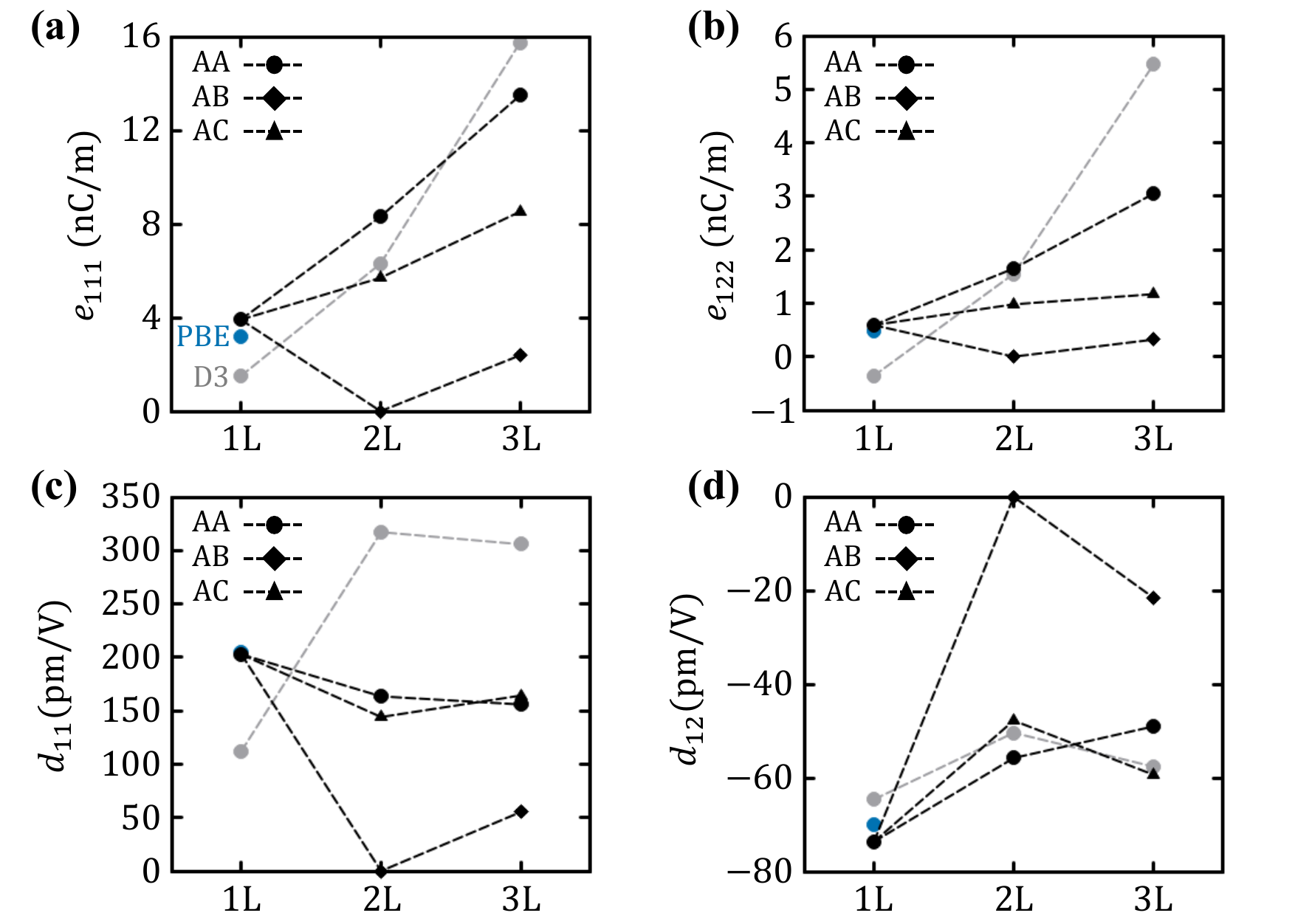}
\caption{2D piezoelectric strain coefficients (a) $e_{111}$ and (b) $e_{122}$, and piezoelectric stress coefficients (c) $d_{11}$ and (d) $d_{12}$) of 1L, 2L, and 3L SnSe with AA, AB, and AC stacking configurations, calculated with the rev-vdW-DF2 vdW correction (black). 
For comparison, the results of the Grimme-D3 corrections (grey) and PBE (blue) are also shown for the AA SnSe and 1L SnSe, respectively.
\label{fig4}}
\end{figure}

Figure~\ref{fig4} shows the variation of four available piezoelectric coefficients ($e_{111}$, $e_{122}$, $d_{11}$, and $d_{12}$) of SnSe with the number of layers up to 3L in AA, AB, and AC stacking configurations. 
(See Table~S3 in SI for more details) 
For the AB stacking configuration, due to its AFE order, the piezoelectric coefficients of the even-numbered layers are calculated to be zero, and those of the odd-numbered layers are also smaller than those of the monolayer.
On the other hand, the 2D piezoelectric strain coefficients ($e_{111}$ and $e_{122}$, in the unit of nC/m) increase with the number of layers in both AA and AC stacking configurations. 
Moreover, AA stacking exhibits higher piezoelectric coefficients than AC stacking, which is consistent with our prediction based on the LCs.
For example, the $e_{111}$ of 3L AA SnSe was calculated to be 13.5~nC/m which is larger than that of AC SnSe (8.53~nC/m), and also six times larger than that of 3L AB SnSe (2.39~nC/m).
Note that the effective bulk piezoelectric coefficient (C/m$^2$) of 3L AA SnSe is also 15\% larger than that of 1L SnSe thanks to the smaller LC.

The other important coefficients, the piezoelectric stress coefficients ($d_{11}$ and $d_{12}$) of multilayer AA SnSe are comparable or slightly larger than those of 1L SnSe, depending on the choice of vdW correction. 
For all other AA MXs, we calculated $e_{111}$, $e_{122}$, $d_{11}$, and $d_{12}$ up to 2L using both rev-vdW-DF2 and Grimme-D3 corrections, as summarized in Tables S4-S6 in SI. 
We consistently found that as the LC decreases, $e_{111}$ and $e_{122}$ continue to enhance, while $d_{11}$ and $d_{12}$ remain comparable to those of 1L MXs.
It is worth noting that although there is no meaningful improvement in $d_{11}$ and $d_{12}$ in multilayer AA MXs compared to 1L, these values are much larger than those of AB MXs. 
For example, the $d_{11}$ of 3L AA SnSe (155.96~pm/V) is almost three times larger than that of 3L AB SnSe (55.2~pm/V), suggesting that AA stacking MX is a promising candidate in piezoelectric applications.

As shown in Fig.~\ref{fig3}(c), the LC $a$ of AA SnSe continues to shrink with stacking, but the bulk AA SnSe loses $P$ and becomes paraelectric, implying that there exists a critical thickness below which non-zero $P$ is accommodated.
In fact, a previous experiment has successfully grown AA SnS on mica substrates, exhibiting room temperature ferroelectricity up to 15 layers,~\cite{higashitarumizu2020purely} indicating that the critical thickness is not too thin and could be further optimized by adjusting the experimental conditions. 
This observation supports our prediction of optimizing the piezoelectric coefficients of multilayer MXs by utilizing AA stacking configuration.

\begin{figure}[t]
\includegraphics[width=1.0\columnwidth]{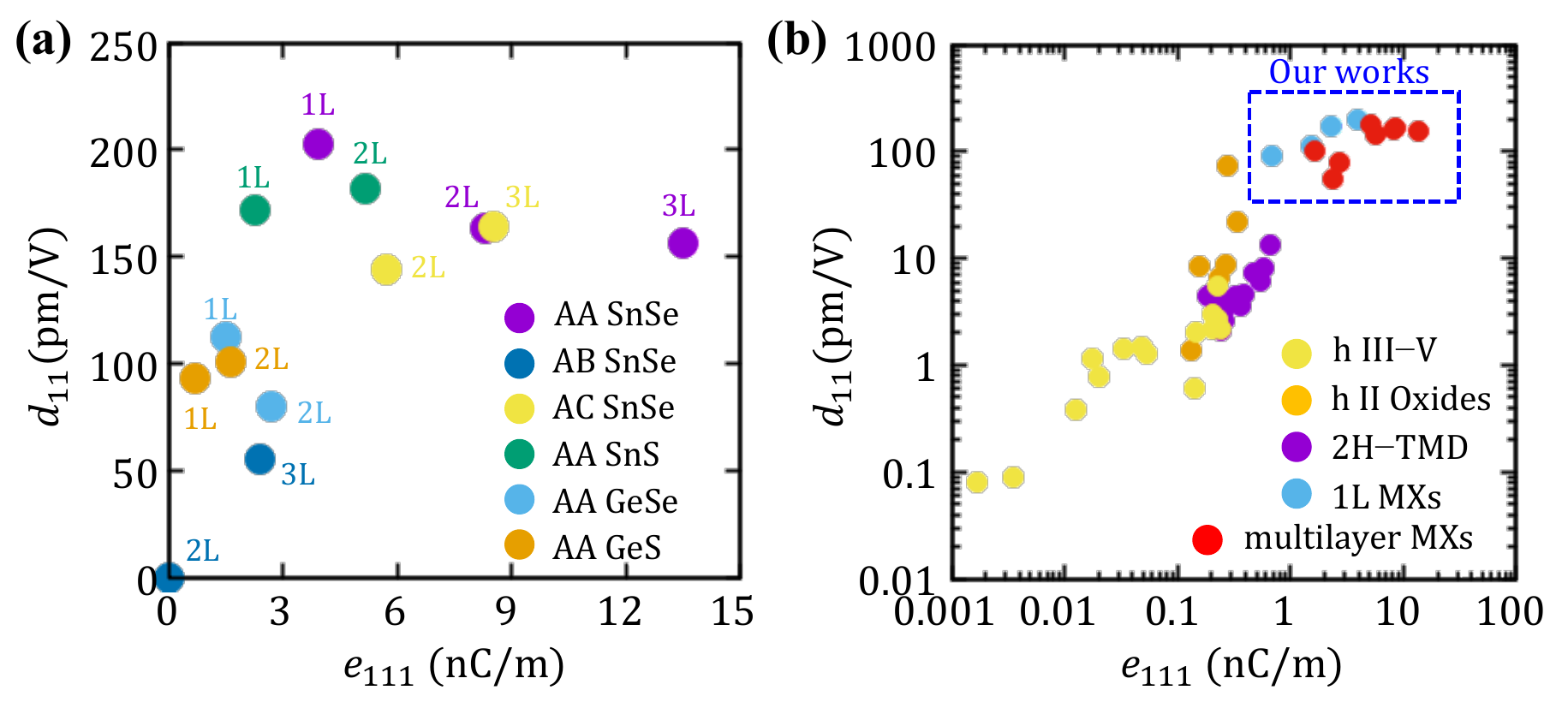}
\caption{Summarized piezoelectric coefficients of (a) group IV monochalcogenides (MXs) and (b) various 2D materials. 
In (a), purple, dark-blue, yellow, green, skyblue, and orange circles represent AA (stacking) SnSe, AB SnSe, AC SnSe, AA SnS, AA GeSe, and AA GeS, respectively. 
In (b), yellow, orange, and purple circles denote materials of hexagonal Group III-V (h III-V), hexagonal Group II oxides (h II Oxides), and H phase transition metal dichalcogenides (2H-TMD), respectively, obtained from Ref. [\!\!\citenum{blonsky2015ab}]. Skyblue and red circles surrounded by a blue dashed rectangle denote our results of 1L MXs and multilayer MXs.
\label{fig5}}
\end{figure}

Finally, we summarize the piezoelectric coefficients of multilayer MXs, as shown in Fig.~\ref{fig5}(a).
SnSe shows the highest $e_{111}$ and $d_{11}$, followed by SnS, GeSe, and GeS. 
In all MXs, the AA stacking boasts higher $e_{111}$ than any other stacking, while $d_{11}$ is more or less than that of the corresponding 1L case regardless of the stacking configuration.
In addition to $e_{111}$ and $d_{11}$, other physical properties such as stability and switching barrier between the FE and AFE phases are also important in piezoelectric applications. Therefore, all MXs can be practically useful, and utilizing AA stacking configuration is an efficient way to optimize their piezoelectric response.

For comparison, we also summarized the piezoelectric coefficients of several other 2D materials, as shown in Fig.~\ref{fig5}(b).
Consistent with the previous reports, we observed that 1L MXs have exceptionally large piezoelectric coefficients compared to the other 2D materials, even larger coefficients in AA stacked MXs.
It is worthy of note that the giant piezoelectric coefficients observed in AA MXs not only stand out among 2D materials, but are also comparable to bulk piezoelectric materials. 
Recent high-throughput DFT calculations revealed that among 941 bulk piezoelectric materials, only 5\% of materials satisfy $|e_{ij}|>3~$C/m$^2$, including experimentally-confirmed giant piezoelectric materials such as BaTiO$_3$ (3.49~C/m$^2$), SrHfO$_3$ (8.73~C/m$^2$), RbTaO$_3$ (8.93~C/m$^2$), and BaNiO$_3$ (27.46~C/m$^2$).~\cite{de2015database}
Using the effective thickness of 1.74~nm (1.74~nm=1.5$c$, $c$ is a out-of-plane lattice constant of bulk SnSe), the effective bulk $e_{111}$ value of 3L AA SnSe was calculated to be 7.76~C/m$^2$, which is even larger than that of BaTiO$_3$.
Within the Grimme-D2 vdW correction, which provides the smallest LCs among all vdW functionals, the $e_{111}$ of 2L AA SnSe was estimated to be 12.66~C/m$^2$ (14.68~nC/m). It is even larger than not only that of 3L AA SnSe estimated by rev-vdW-DF2, but also that of SrHfO$_3$ and RbTaO$_3$.
We now conclude that our theoretical calculations have undoubtedly revealed that there is a strong correlation between the reduction of LCs and giant piezoelectric coefficients in AA-stacked MXs, regardless of the vdW functional used, making it a promising candidate for low-dimensional piezoelectric applications.

\section{Conclusion}
\label{Conclusion}

In summary, we have used first-principles calculations to study the piezoelectric coefficients of multilayer MXs with AA and AC stacking configurations, which are known to stabilize ferroelectric order.
Our re-examination of 1L MXs revealed the pivotal role of the van der Waals interaction in ensuring accurate and reliable predictions of piezoelectric coefficients.
Through systematic DFT calculations, we found that the AA-stacked configurations exhibit remarkably larger piezoelectric coefficients compared to all reported layered materials, including their monolayer counterparts.
The origin of this enhancement lies in the compressive strain along the armchair direction, which is spontaneously introduced in the ferroelectric AA stacking. These findings not only provide a comprehensive understanding of the piezoelectric behavior of MXs, but also suggest a novel strategy for optimizing piezoelectricity in low-dimensional materials through stacking configuration.

\acknowledgments
S.L., W.J., and T.L. were supported by the National Science Foundation (NSF) through the DMREF program under Award No. DMR-1921629.
S.L. is also supported by Basic Science Research Program through the National Research Foundation of Korea (NRF) funded by the Ministry of Education (NRF-2021R1A6A3A14038837).
H.-R.K. and Y.-K.K. acknowledge financial support from the Korean government through the NRF (NRF-2022R1A2C1005505, NRF2022M3F3A2A01073562).
The computational work was partially done using the resources of the KISTI Supercomputing Center (KSC-2023-CRE-0053).

\end{document}



Supplementary Information \\

\title{Giant piezoelectricity in group IV monochalcogenides with ferroelectric AA layer stacking}

\author{Seungjun Lee}
\affiliation{Department of Electrical and Computer Engineering, University of Minnesota, Minneapolis, Minnesota 55455, USA}

\author{Hyeong-Ryul Kim}
\affiliation{Department of Physics and Research Institute for Basic Sciences, Kyung Hee University, Seoul 02447, Korea}

\author{Wei Jiang}
\affiliation{Department of Electrical and Computer Engineering, University of Minnesota, Minneapolis, Minnesota 55455, USA}

\author{Young-Kyun Kwon}\email{ykkwon@khu.ac.kr}
\affiliation{Department of Physics and Research Institute for Basic Sciences, Kyung Hee University, Seoul 02447, Korea}
\affiliation{Department of Information Display, Kyung Hee University, Seoul 02447, Korea}

\author{Tony Low}\email{tlow@umn.edu}
\affiliation{Department of Electrical and Computer Engineering, University of Minnesota, Minneapolis, Minnesota 55455, USA}

\date{\today}
\maketitle

\begin{figure}[h]
\includegraphics[width=1.00\columnwidth]{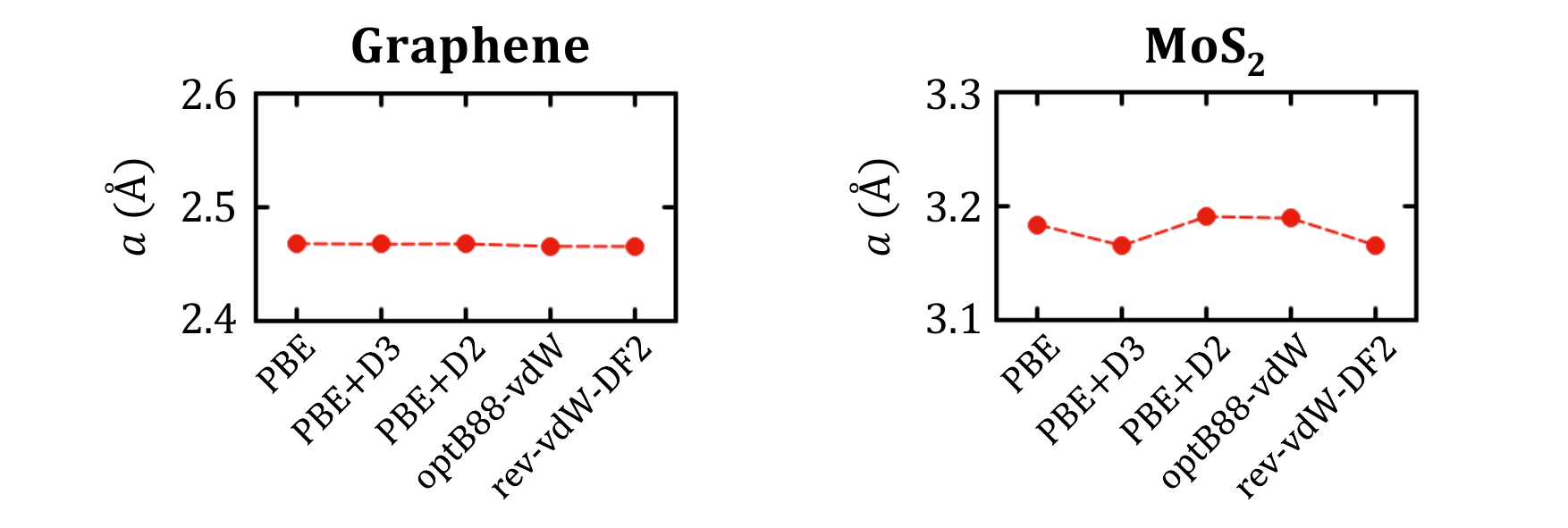}
\caption{Calculated lattice constants of graphene and monoalyer H-MoS$_2$, obtained using different exchange-correlation functionals.
\label{figs1}}
\end{figure}

\clearpage
\newpage

\begin{figure}[h]
\includegraphics[width=1.00\columnwidth]{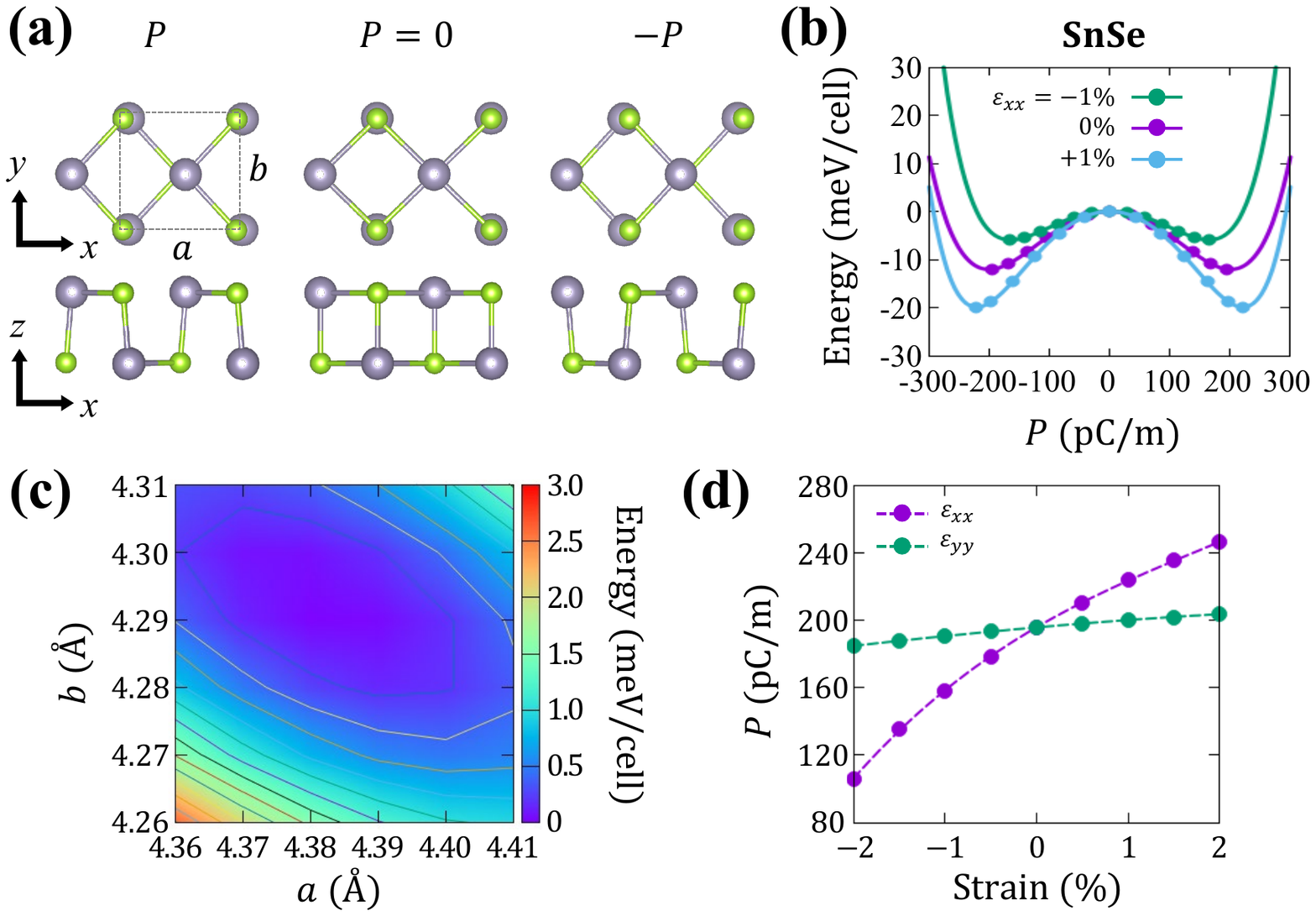}
\caption{Computational procedure in determining piezoelectric coefficient of 1L SnSe.
(a) Two mirror symmetric ground states of 1L SnSe (${P}$ and $-{P}$) and corresponding inversion symmetric structures with ${P}=0$ between their reaction path. 
(b) Calculated spontaneous polarization ${P}$ along with the reaction path between ${P}$ and $-{P}$ states. The energy barrier was evaluated using the nudged elastic band (NEB) calculation method.~\cite{henkelman2000climbing}
As discussed in the main manuscript, the compressive (tensile) strain along the armchair direction decreases (increases) its ${P}$.
(c) Total energy contour map of 1L SnSe as a function of lattice constants. We calculated elastic coefficients of $C_{ij}$ using Eq.~(1) in the main manuscripts. 
(d) Calculated ${P}$ of 1L SnSe as function of uniaxial strain. Dots indicate DFT results, and the dashed lines show the polynomial fitting function. Piezoelectric strain coefficient $e_{ijk}$ was evaluated through Eq.~(2) in the main manuscripts, and piezoelectric stress coefficient $d_{ij}$ was evaluated through Eq.~(3, 4) using obtained $C_{ij}$ and $e_{ijk}$.
In this example, all calculation results were obtained within PBE XC functional. 
Similar calculations were repeated to obtain piezoelectric coefficients of other materials, multilayer cases, as well as the same material with different van der Waals functionals.
\label{figs2}}
\end{figure}

\clearpage
\newpage

\begin{figure}[h]
\includegraphics[width=1.00\columnwidth]{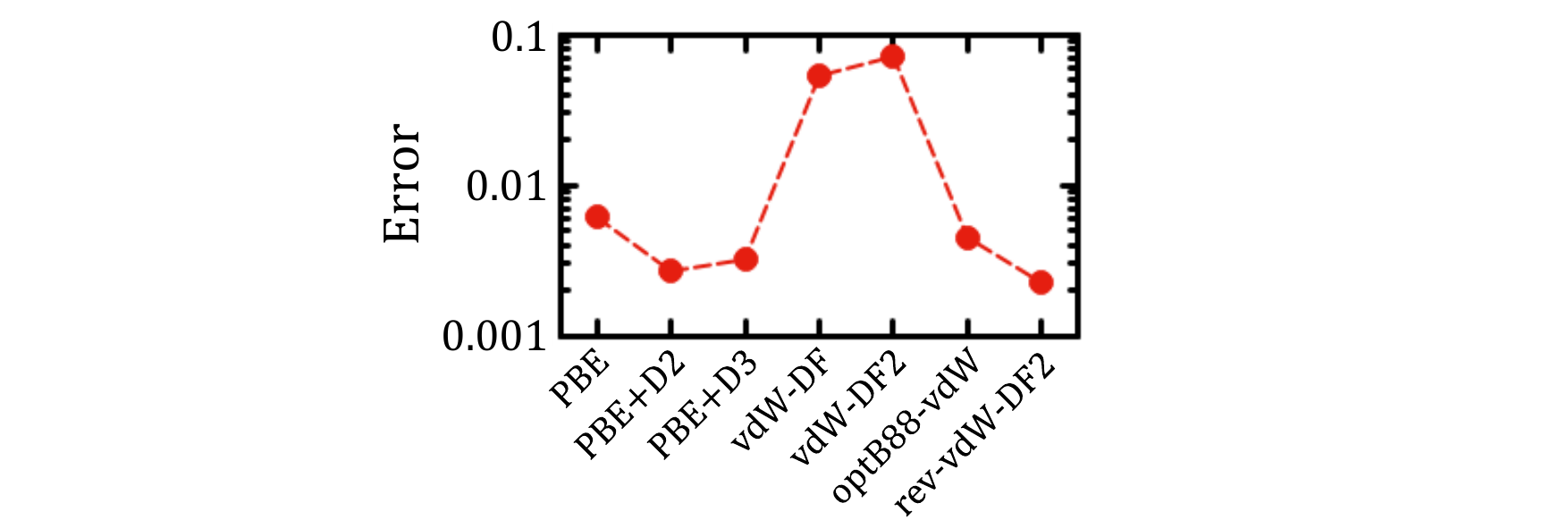}
\caption{Performance of van der Waals functionals for bulk group IV monochalcogenides.
Error is evaluated by $\sum_{ij}$$((a^{i,j}_{\textrm{thy}}-a^{i,j}_{\textrm{exp}})/a^{i,j}_{\textrm{exp}})^2$, where 
$a_{\textrm{thy}}$ and $a_{\textrm{exp}}$ are the lattice constants obtained from DFT calculation and experiment, $i$ is a index for the three lattice constants ($a, b,$ and $c$ in Table~\ref{Tables1}) and $j$ is a index for the four materials studied in this work (SnSe, SnS, GeSe, and GeS).
\label{figs3}}
\end{figure}

\clearpage
\newpage

\begin{table}
\caption {Calculated lattice constants of bulk group IV monochalcogenides, obtained using various van der Waals corrections. Experimental values are also shown for comparison.
\label{table1}}
\begin{ruledtabular}
\begin{tabular}{ llll llll} 
 bulk-SnSe  & ${\textit{a}}$ (\AA) & ${\textit{b}}$ (\AA) & ${\textit{c}}$ (\AA) & bulk-SnS & ${\textit{a}}$ (\AA) & ${\textit{b}}$ (\AA) & ${\textit{c}}$ (\AA) \\
 \cline{1-4}\cline{5-8}
PBE	&	4.546	&	4.207	&	11.762	&	PBE	&	4.426	&	4.023	&	11.415	\\
D2	&	4.349	&	4.197	&	11.631	&	D2	&	4.261	&	4.012	&	11.384	\\
D3	&	4.521	&	4.166	&	11.553	&	D3	&	4.409	&	3.990	&	11.184	\\
vdW-DF	&	4.791	&	4.263	&	12.295	&	vdW-DF	&	4.647	&	4.073	&	11.920	\\
vdW-DF2	&	4.857	&	4.334	&	12.512	&	vdW-DF2	&	4.703	&	4.133	&	12.076	\\
optB88	&	4.501	&	4.22	&	11.761	&	optB88	&	4.365	&	4.039	&	11.400	\\
rev-vdW-DF2	&	4.429	&	4.201	&	11.627	&	rev-vdW-DF2	&	4.291	&	4.025	&	11.282	\\
Experiment\cite{chattopadhyay1986neutron}	&	4.445	&	4.153	&	11.501	&	Experiment\cite{chattopadhyay1986neutron}	&	4.34	&	3.97	&	11.14	\\
 \hline
bulk-GeSe  & ${\textit{a}}$ (\AA) & ${\textit{b}}$ (\AA) & ${\textit{c}}$ (\AA) & bulk-GeS & ${\textit{a}}$ (\AA) & ${\textit{b}}$ (\AA) & ${\textit{c}}$ (\AA) \\
 \cline{1-4}\cline{5-8}
 PBE	&	4.514	&	3.892	&	11.179	&	PBE	&	4.428	&	3.679	&	10.757	\\
D2	&	4.469	&	3.849	&	11.011	&	D2	&	4.339	&	3.670	&	10.660	\\
D3	&	4.513	&	3.877	&	10.978	&	D3	&	4.452	&	3.671	&	10.634	\\
vdW-DF	&	4.779	&	3.942	&	11.802	&	vdW-DF	&	4.664	&	3.727	&	11.337	\\
vdW-DF2	&	4.782	&	4.027	&	11.924	&	vdW-DF2	&	4.669	&	3.795	&	11.388	\\
optB88	&	4.443	&	3.927	&	11.158	&	optB88	&	4.338	&	3.720	&	10.712	\\
rev-vdW-DF2	&	4.339	&	3.915	&	10.977	&	rev-vdW-DF2	&	4.229	&	3.713	&	10.544	\\
Experiment\cite{Murgatroyd2020}	&	4.395	&	3.836	&	10.833	&	Experiment\cite{PhysRevB.16.832}	&	4.30	&	3.64	&	10.47	\\
\end{tabular}
\end{ruledtabular}
\label{Tables1}
\end{table}


\clearpage
\newpage

\begin{table}
\caption {Calculated lattice constants of multilayer group IV monochalcogenides, obtained using various van der Waals corrections.
\label{table1}}
\begin{ruledtabular}
\begin{tabular}{ llllll } 
 Materials  & vdW & Stacking & Thickness & ${\textit{a}}$ (\AA) & ${\textit{b}}$ (\AA)  \\
  \hline
SnSe  	&	rev-vdW-DF2	&	2L	&	AA	&	4.302	&	4.263	\\
	&		&		&	AB	&	4.392	&	4.230	\\
	&		&		&	AC	&	4.340	&	4.251	\\
 \cline{3-6}
	&		&	3L	&	AA	&	4.290	&	4.257	\\
	&		&		&	AB	&	4.394	&	4.224	\\
	&		&		&	AC	&	4.339	&	4.248	\\
\cline{3-6}
	&		&	bulk	&	AA	&	4.257	&	4.257	\\
	&		&		&	AB	&	4.429	&	4.201	\\
	&		&		&	AC	&	4.339	&	4.238	\\
 \cline{2-6}
 	&	D3	&	2L	&	AA	&	4.351	&	4.265	\\
	&		&		&	AC	&	4.464	&	4.251	\\
 \cline{2-6}
 &	D2	&	2L	&	AA	&	4.251	&	4.229	\\
&		&		&	AC	&	4.334	&	4.215	\\
\cline{2-6}
&	optB88	&	2L	&	AA	&	4.356	&	4.281	\\
&		&		&	AC	&	4.398	&	4.27	\\
\cline{1-6}
SnS	&	rev-vdW-DF2	&	2L	&	AA	&	4.170	&	4.071	\\
	&		&		&	AC	&	4.195	&	4.052	\\
 \cline{1-6}
GeSe	&	rev-vdW-DF2	&	2L	&	AA	&	4.150	&	3.972	\\
	&		&		&	AC	&	4.254	&	3.961	\\
 \cline{1-6}
GeS	&	rev-vdW-DF2	&	2L	&	AA	&	4.089	&	3.780	\\
	&		&		&	AC	&	4.241	&	3.721	\\
 \end{tabular}
\end{ruledtabular}
\label{Tables2}
\end{table}

\clearpage
\newpage

\begin{table*}[h]
\caption {Lattice constants, $a$ and $b$, spontaneous polarization, $P$, piezoelectric strain coefficients, $e_{ijk}$, planar elastic stiffness coefficients, $C_{ij}$, and piezoelectric stress coefficients, $d_{ij}$ of multilayer SnSe calculated by various functionals.
\label{table1}}
\begin{ruledtabular}
\begin{tabular}{c c c c c c c c c c c c } 
Layer & vdW & \makecell{$a$ \\ (\AA)} & \makecell{$b$ \\ (\AA)} &\makecell{${P}$ \\ (pC/m)} & \makecell{$e_{111}$  \\ (nC/m)} & \makecell{$e_{122}$  \\ (nC/m)} & \makecell{$C_{11}$  \\ (N/m)} & \makecell{$C_{22}$  \\ (N/m)} & \makecell{$C_{12}$  \\ (N/m)} & \makecell{$d_{11}$  \\ (pm/V)} & \makecell{$d_{12}$  \\ (pm/V)} \\
\hline
2L AA 	&	ref-vdW-DF2	&	4.302 & 4.263	&	249.1	&	8.32	&	1.65	&	65.69	&	97.67	&	43.41	&	163.52	&	-55.78	\\
	&	D3	&	4.351 & 4.265	&	398.0	&	6.31	&	1.55	&	22.45	&	70.56	&	16.09	&	317.24	&	-50.38	\\
	&	D2	&	4.251 & 4.229	&	146.0	&	14.68	&	5.41	&	46.97	&	87.19	&	37.01	&	396.15	&	-155.67	\\
	&	optB88	&	4.356 & 4.281	&	336.1	&	6.02	&	0.9	&	56.51	&	96.68	&	40.96	&	144	&	-51.7	\\
 \hline
3L AA	&	ref-vdW-DF2	&	4.290 & 4.257	&	347.5	&	13.5	&	3.05	&	107.82	&	152.92	&	67.63	&	155.96	&	-49.03	\\
	&	D3	&	4.308 & 4.256	&	498.7	&	15.76	&	5.47	&	59.18	&	121.46	&	40.74	&	305.95	&	-57.59	\\
 \hline
2L AC	&	ref-vdW-DF2	&	4.340 & 4.251	&	381.9	&	5.7	&	0.98	&	51.89	&	92.05	&	37.29	&	144.17	&	-47.76	\\
	&	D3	&	4.464 & 4.251	&	556.2	&	2.54	&	-0.38	&	31.89	&	80.25	&	26.75	&	116.08	&	-43.43	\\
 \hline
3L AC	&	ref-vdW-DF2	&	4.339 & 4.248	&	605.1	&	8.53	&	1.17	&	72.58	&	138.38	&	57.12	&	164.22	&	-59.33	\\
\hline
2L AB	&	ref-vdW-DF2	&	4.392 & 4.230	&		&		&		&		&		&		&		&		\\
\hline
3L AB	&	ref-vdW-DF2	&	4.394 & 4.224	&	230.9	&	2.39	&	0.32	&	65.26	&	129.79	&	56.37	&	55.2	&	-21.51	\\
\end{tabular}
\end{ruledtabular}
\label{Tables3}
\end{table*}

\clearpage
\newpage

\begin{table*}[h]
\caption {Lattice constants, $a$ and $b$, spontaneous polarization, $P$, piezoelectric strain coefficients, $e_{ijk}$, planar elastic stiffness coefficients, $C_{ij}$, and piezoelectric stress coefficients, $d_{ij}$ of 1L SnS, GeSe, and GeS calculated by PBE functional.
\label{table1}}
\begin{ruledtabular}
\begin{tabular}{c c c c c c c c c c c c } 
Material & layer & \makecell{$a$ \\ (\AA)} & \makecell{$b$ \\ (\AA)} & \makecell{${P}$ \\ (pC/m)} & \makecell{$e_{111}$  \\ (nC/m)} & \makecell{$e_{122}$  \\ (nC/m)} & \makecell{$C_{11}$  \\ (N/m)} & \makecell{$C_{22}$  \\ (N/m)} & \makecell{$C_{12}$  \\ (N/m)} & \makecell{$d_{11}$  \\ (pm/V)} & \makecell{$d_{12}$  \\ (pm/V)} \\
\hline
SnS	&	1L	&	4.284 & 4.082	&	256.2	&	1.8	&	-0.13	&	22.33	&	38.97	&	15.98	&	117.47	&	-51.5	\\
GeSe	&	1L	&	4.275 & 3.981	&	330.6	&	1.01	&	-0.62	&	19.43	&	52.92	&	17.47	&	88.9	&	-41.06	\\
GeS	&	1L	&	4.487 & 3.660	&	448.5	&	0.41	&	-0.81	&	11.62	&	52.52	&	18.5	&	136.25	&	-63.41	\\
\end{tabular}
\end{ruledtabular}
\label{Tables4}
\end{table*}

\begin{table*}[h]
\caption {Lattice constants, $a$ and $b$, spontaneous polarization, $P$, piezoelectric strain coefficients, $e_{ijk}$, planar elastic stiffness coefficients, $C_{ij}$, and piezoelectric stress coefficients, $d_{ij}$ of 1L and AA-stacked 2L SnS, GeSe, and GeS calculated by PBE with ref-vdW-DF2 functional.
\label{table1}}
\begin{ruledtabular}
\begin{tabular}{c c c c c c c c c c c c } 
Material & layer & \makecell{$a$ \\ (\AA)} & \makecell{$b$ \\ (\AA)} & \makecell{${P}$ \\ (pC/m)} & \makecell{$e_{111}$  \\ (nC/m)} & \makecell{$e_{122}$  \\ (nC/m)} & \makecell{$C_{11}$  \\ (N/m)} & \makecell{$C_{22}$  \\ (N/m)} & \makecell{$C_{12}$  \\ (N/m)} & \makecell{$d_{11}$  \\ (pm/V)} & \makecell{$d_{12}$  \\ (pm/V)} \\
\hline
SnS	&	1L	&	4.175 & 4.056	&	210.2	&	2.27	&	-0.14	&	18.68	&	39.21	&	14.27	&	172.09	&	-66.20	\\
	&	2L AA	&	4.170 & 4.071	&	384.2	&	5.15	&	-0.58	&	42.88	&	77.82	&	32.07	&	181.67	&	-82.32	\\
 \hline
GeSe	&	1L	&	4.201 & 3.985	&	274.1	&	1.51	&	-0.46	&	23.03	&	53.78	&	20.70	&	112.00	&	-51.66	\\
	&	2L AA	&	4.150 & 3.972	&	597.0	&	2.71	&	-1.09	&	56.15	&	114.36	&	44.07	&	79.92	&	-40.33	\\
 \hline
GeS	&	1L	&	4.253 & 3.722	&	395.5	&	0.68	&	-0.81	&	21.58	&	54.50	&	23.85	&	92.84	&	-55.49	\\
	&	2L AA	&	4.089 & 3.780	&	685.1	&	1.64	&	-1.55	&	44.04	&	110.96	&	48.38	&	100.93	&	-57.97	\\
\end{tabular}
\end{ruledtabular}
\label{Tables5}
\end{table*}

\clearpage
\newpage

\begin{table*}[h]
\caption {Lattice constants, $a$ and $b$, spontaneous polarization, $P$, piezoelectric strain coefficients, $e_{ijk}$, planar elastic stiffness coefficients, $C_{ij}$, and piezoelectric stress coefficients, $d_{ij}$ of 1L and AA-stacked 2L SnS, GeSe, and GeS calculated by PBE with Grimme-D3 functional.
\label{table1}}
\begin{ruledtabular}
\begin{tabular}{c c c c c c c c c c c c } 
Material & layer & \makecell{$a$ \\ (\AA)} & \makecell{$b$ \\ (\AA)} & \makecell{${P}$ \\ (pC/m)} & \makecell{$e_{111}$  \\ (nC/m)} & \makecell{$e_{122}$  \\ (nC/m)} & \makecell{$C_{11}$  \\ (N/m)} & \makecell{$C_{22}$  \\ (N/m)} & \makecell{$C_{12}$  \\ (N/m)} & \makecell{$d_{11}$  \\ (pm/V)} & \makecell{$d_{12}$  \\ (pm/V)} \\
\hline
SnS	&	1L	&	 4.441 & 4.032	&	327.4	&	0.99	&	-0.35	&	14.8	&	41.62	&	16.53	&	137.03	&	-62.82	\\
	&	2L AA	&	4.455 & 4.03	&	653.3	&	2.01	&	-0.87	&	24.18	&	74.04	&	27.61	&	168.14	&	-74.45	\\
 \hline
GeSe	&	1L	&	4.266 & 3.978	&	334.7	&	0.95	&	-0.66	&	22.8	&	57.03	&	23.14	&	90.81	&	-48.41	\\
	&	2L AA	&	4.255 & 3.967	&	724.5	&	1.49	&	-1.52	&	62.95	&	122.98	&	55.02	&	56.61	&	-37.69	\\
 \hline
GeS	&	1L	&	 4.530 & 3.631	&	461.3	&	0.34	&	-0.81	&	10.2	&	55	&	18.18	&	145.02	&	-62.66	\\
	&	2L AA	&	4.428 & 3.656	&	856.3	&	0.69	&	-1.65	&	35.9	&	119.75	&	39.53	&	54.03	&	-31.61	\\
\end{tabular}
\end{ruledtabular}
\label{Tables6}
\end{table*}
